\documentclass[pra,letterpaper,aps,10pt,superscriptaddress,twocolumn,floatfix,showpacs]{revtex4-1}
\usepackage{amsmath,graphicx,amssymb,braket,color,subfigure,upgreek}
\usepackage{units}
\usepackage{scalerel,stackengine}

\usepackage{amsfonts}
\usepackage{epsfig}
\usepackage{epstopdf}
\DeclareGraphicsExtensions{.pdf,.eps,.png,.jpg,.mps}
\usepackage{microtype}
\usepackage{bbm}
\usepackage{amsmath,amssymb,amsfonts,latexsym,dcolumn,bm,amsbsy}
\usepackage[colorlinks, linkcolor=blue, citecolor=blue, urlcolor=blue, breaklinks=true]{hyperref}
\usepackage[english]{babel}
\usepackage[utf8]{inputenc}
\usepackage{textcomp}
\begin{document}

\title{Prospects of reinforcement learning for the simultaneous damping of many mechanical modes}
\author{Christian Sommer}
\affiliation{Max Planck Institute for the Science of Light, Staudtstra{\ss}e 2,
D-91058 Erlangen, Germany}
\author{Muhammad Asjad}
\affiliation{Max Planck Institute for the Science of Light, Staudtstra{\ss}e 2,
D-91058 Erlangen, Germany}
\author{Claudiu Genes}
\affiliation{Max Planck Institute for the Science of Light, Staudtstra{\ss}e 2,
D-91058 Erlangen, Germany}
\affiliation{Department of Physics, University of Erlangen-Nuremberg, Staudtstra{\ss}e 2,
D-91058 Erlangen, Germany}
\date{\today}


\begin{abstract}
We apply adaptive feedback for the partial refrigeration of a mechanical resonator, i.e. with the aim to simultaneously cool the classical thermal motion of more than one vibrational degree of freedom. The feedback is obtained from a neural network parametrized policy trained via a reinforcement learning strategy to choose the correct sequence of actions from a finite set in order to simultaneously reduce the energy of many modes of vibration. The actions are realized either as optical modulations of the spring constants in the so-called quadratic optomechanical coupling regime or as radiation pressure induced momentum kicks in the linear coupling regime. As a proof of principle we numerically illustrate efficient simultaneous cooling of four independent modes with an overall strong reduction of the total system temperature.
\end{abstract}

\maketitle

The radiation pressure effect of light onto the motion of mechanical resonators has been extensively employed to bring such macroscopic systems towards the quantum ground state~\cite{Aspelmeyer2014cavity,windey2019cavity,deli2019cavity, Rossi2017, Clark2017, Qiu2019,Asenbaum2013,Mancini1998, Clemens2016, Kiesel2013,Millen2015}.
In a standard approach, the aim is to isolate a single vibrational mode and bring it to a state where the only relevant motion is given by the zero-point fluctuations. Cold-damping is one of the used techniques, where one detects motionally-induced phase changes in the cavity output and an electronic feedback loop is implemented to dynamically modify the cavity drive such as to produce an extra optical damping effect~\cite{Genes2008Ground, Steixner2005, Bushev2006,Rossi2018, Cohadon1999, Poggio2007, Wilson2015,Tebbenjohanns2018Cold}. Alternatively, in the good cavity limit where the photon loss rate is smaller than the mechanical frequency, the resolved sideband technique can be implemented by detuning the drive to the cooling sideband~\cite{Gigan2006self, Braginsky2007parametric, Marquardt2007Quantum, Wilson2007Theory, Teufel2011Sideband}. As the effect stems from the inherent time delay between the action of the mechanical resonator onto the cavity field and the back-action of light, this can be seen as a sort of automatic cavity induced feedback. Both techniques are devised and have been successfully applied for single vibrational mode cooling. However, it is interesting to devise an alternative technique that can induce partial to full refrigeration of the mechanical resonator, i.e. to simultaneously cool a multitude of vibrational modes into which the thermal energy is distributed. An impediment is that the detected output signal only gives information on a generalized collective quadrature but not on all modes. This leads to efficient cooling of some collective mode (for example center of mass) while some collective modes become dark and remain in a high temperature state. It has been recently pointed out that some strategies such as multimode cold-damping could in principle lead to sympathetic cooling of many modes via disorder induced coupling between bright and dark modes~\cite{sommer2019partial}.\\
\indent Here, we propose a machine learning approach towards devising a strategy capable of providing refrigeration of the classical motion of a mechanical resonator based on the feedback obtained from the detection of a single optical mode. While the detected optical mode only gives information on a collective generalized quadrature obtained as a linear combination of individual mode displacements, the procedure is optimized such as at any instant in time a compromise is made between efficiently cooling a particular target mode while not affecting the others too much. We provide proof-of-principle multi-mode numerical simulations using a neural network parametrized policy trained by a reinforcement learning algorithm to generate the feedback signal capable of simultaneously extracting thermal energy from four distinct modes of a single mechanical resonator.\\
\indent Machine learning techniques have been recently applied to various applications in quantum physics ranging from the identification of phases in many-body systems, predicting ground-state energies for electrostatic potentials, active learning approaches to propose and optimize experimental setup configurations and towards applications for quantum control and quantum-error correction~\cite{Chen2014Fidelity, Carrasquilla2017Machine, Nieuwenburg2017Learning, Dunjko2017Machine, Carleo2017Solving, Mills2017Deep, Melnikov2018Active, Bukov2018Reinforce, Foesel2018Reinforce}. In particular, a few studies~\cite{Chen2014Fidelity, Foesel2018Reinforce, sweke2018reinforcement} successfully applied the technique of reinforcement learning with neural networks~\cite{Russel2018Modern}. This approach originates from the idea, to let an intelligent agent that observes its environment choose an action, that is determined by a given policy trying to optimize a particular reward and/or minimize a punishment.\\
\indent We employ such a reinforcement learning technique for optically assisted cooling of the classical thermal state of a multi-mode mechanical resonator system~\cite{Nielsen2017Multimode, Piergentili2018, PhysRevA.99.023851}. The learning technique allows one to acquire a nonlinear function that chooses a feedback action that will be applied on the dynamical system upon taking the full or partial measured state of the system as an input. The training of this function that is given by a dense neural network is obtain by trial and error and quantified by an increased reward that is obtained by successfully reducing the energy of the resonators.\\
\begin{figure*}[t]
\includegraphics[width=1.96\columnwidth]{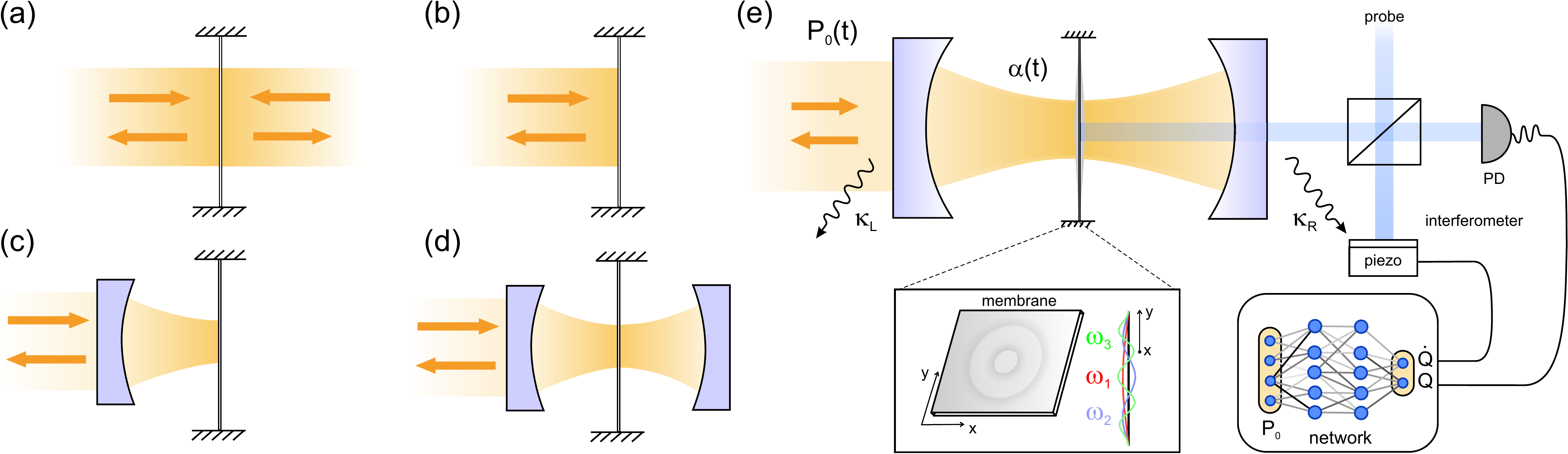}
\centering
\caption{Cooling of thermal motion via two-sided kicking in (a) or via one sided-kicking in (b). (c) Increased damping efficiency can be realized by an amplification of the photon-phonon coupling in the linear coupling regime via an optical cavity enhancement of the electric field amplitude. (d) Membrane-in-the-middle configuration leads to a quadratic coupling in the mechanical displacement allowing optical control of the mechanical mode's spring constant. (e) Simultaneous cooling of multiple oscillating modes (inset shows a few drum modes of a dielectric membrane) using feedback generated by a reinforcement trained policy, encoded in a neural network. While the illustration shows a quadratic membrane-in-the-middle setup, the validity extends to the end-mirror linear setup as well. The outgoing signal from a driven cavity carries information on the collective displacement of all membrane vibration modes. This signal is fed through a neural network and the network's suggested action is implemented as a modulation of the cavity input drive amplitude.}
\label{fig1}
\end{figure*}
\indent The physical systems considered are depicted in Fig.~\ref{fig1}. The mechanical resonator is subject to environmental noise described by a standard Brownian motion stochastic force leading to thermalization at some equilibrium temperature $T$. The feedback action is implemented via the radiation pressure force, i.e. photon kicks either from one or two sides. The induced damping is straightforward in the two-sided kicking case [illustrated in Fig.~\ref{fig1}a]: the read-out of motion is followed by appropriate kicking action from the side towards which the resonator is moving. However, one-sided kicking [illustrated in Fig.~\ref{fig1}b] already suffices allowing setups such as the cavity optomechanical platform pictured in Fig.~\ref{fig1}c. The typical weak free space photon-phonon interaction can also be drastically increased by the filtering of the action through the high-finesse optical cavity. Such a situation is characterized by a linear coupling of the photon number to the membrane's displacement and has been extensively studied in single mode cooling via cavity time delayed effects~\cite{Genes2008Ground} or by implementation of cold damping techniques~\cite{Genes2008Ground} especially in the bad cavity regime. The membrane-in-the-middle \cite{Thompson_2008, Jayich_2008, Asjad2014Robust} scenario in Fig.~\ref{fig1}d,e corresponds to a quadratic coupling in displacement leading to the possibility of optically modulating the mechanical oscillation frequency~\cite{Asjad2014Robust}. We describe in Fig.~\ref{fig1}e a possible approach for feedback cooling via cavity field detection and neural network assisted feedback.\\
\indent We will consider the bad cavity case where losses are large compared to the mechanical resonator's vibration frequencies such that the cavity back-action is negligible. In such a case, the situations described in Fig.~\ref{fig1}b and Fig.~\ref{fig1}c are physically equivalent with the difference that in Fig.~\ref{fig1}c the action of a single photon is multiplied by a large number roughly proportional to the finesse of the cavity. We also distinguish between a parametric regime with quadratic coupling implemented in the membrane-in-the-middle setup and the linear coupling regime realizable with a single-end mirror cavity or in free space. First we analyze the performance of a neural network suggested set of actions onto the cooling of a single mode via parametric modulation of the oscillation frequency: we describe the shape of the action and numerically show the efficient reduction of energy from the initial thermal distribution. We then apply the technique to the linear cooling of four distinct modes of the resonator and find a more complex set of actions required for efficient simultaneous cooling of all four modes (with limitations arising due to the numerical complexity of the simulations).\\

\noindent \textbf{Model} --- We consider a membrane resonator with a few modes of oscillations of frequencies $\omega_j$ (where $j=1,...N$). We start with a quantum formulation of the system's dynamics aimed at future treatments of cooling in the presence of quantum noise. However the current formulation aims only at the reduction of classical thermal noise and is therefore obtained by inferring the equivalent classical stochastic equations of motion. The Hamiltonian for the collection of modes is written as $H_m =\sum_{j=1} \hbar\omega_j/2\left(p_{j}^{2} + q_{j}^{2} \right),$ in terms of dimensionless position and momentum quadratures $q_{j}$ and $p_{j}$ for each independent membrane oscillation mode. The effect of the thermal reservoir can be easily included in a set of equations of motion supplemented with the proper input stochastic noise terms:
\begin{subequations}
\label{FreeEq.1}
\begin{align}
\dot{q}_j &=\omega_{j} p_j, \\
\dot{p}_j &= -\omega_{j}q_j - \gamma_{j} p_j + \xi_j + F_{j}(t).
\end{align}
\end{subequations}
The parameter $\gamma_{j}$ describes the damping of the $j$'s resonator mode. Its associated zero-averaged Gaussian stochastic noise term leading to thermalization with the environment can be fully described by the two-time correlation function:
\begin{eqnarray}
\label{FreeEq.2}
\langle \xi_j(t) \xi_{j'}(t')\rangle &=& \frac{\gamma_{j}}{\omega_j}\int_{0}^{\Omega} \frac{d\omega}{2\pi}e^{-i\omega(t-t')} S_{\text{th}}(\omega)\delta_{jj'},
\end{eqnarray}
where $\Omega$ is the frequency cutoff of the reservoir and the thermal noise spectrum is given by $S_{\text{th}}(\omega)=\omega[\coth\left(\hbar \omega/2k_B T\right) + 1]$. For sufficiently high temperatures $k_B T \gg \hbar \omega_j$, the correlation function becomes a standard white noise input with delta correlations both in frequency and time. Specifically, one can approximate $\langle \xi_j(t) \xi_{j'}(t')\rangle \approx (2\bar{n}_{j}+1)\gamma_{j}\delta(t-t')\delta_{jj'}$, where the occupancy of each vibrational mode is given by $\bar{n}_{j} = (\exp(\hbar \omega_j/k_B T)-1)^{-1} \approx k_B T/\hbar \omega_j$. For numerical simulations we generate a stochastic input noise as a delta-correlated Wiener increment with variance proportional to the integration time-step (see Methods) and follow an approach described in Ref.~\cite{Higham2001An}. For consistency we check (in the Methods) that the thermal bath indeed correctly describes the expected thermalization of an initially cold oscillator towards the equilibrium temperature $T$ at a rate given by $\gamma$.\\
\indent The momentum kicks selected by the network are encompassed in the action of the force terms $F_{j}(t)$. This can be realized for example by the radiation pressure effect of a laser beam, modulated by a device like an AOM (acousto-optic modulator). Here, forces acting on different resonators given by $F_j$ and $F_{j'}$ for $j \neq j'$ differ only by a constant multiplication factor as they are all obtained from the same quantity (the output field).\\
To amplify the effect of the action force onto the mechanical resonator one can utilize optical cavities. A cavity also allows control over the coupling by placing the membrane either in a node (quadratic coupling) or anti-node (linear coupling) of the cavity mode. The Hamiltonian is now modified by the addition of the free cavity mode $\hbar \omega_c a^{\dagger} a$, laser driving resonant to the cavity mode $i\hbar \mathcal{E}(t)\left(a^{\dagger} -a \right)$ (in a frame rotating at $\omega_c$) and optomechanical interaction of linear $\sum_j \hbar g^{(1)}_j a^{\dagger} a q_j$ or quadratic form $\sum_j \hbar g^{(2)}_j a^{\dagger} a q^2_j$. The amplification effect of the light field amplitude can be seen from the relation $\mathcal{E}(t) = \sqrt{2\mathcal{P}_{0}(t)\kappa_L /\hbar \omega_{c}}$ connecting the driving amplitude to the input laser power $\mathcal{P}_{0}(t)$ through the left mirror with losses at rate $\kappa_L$. For high-finesse cavities photons perform many round trips before leaking out through the mirrors resulting in a large momentum transfer onto the mirror: this can be seen by taking the limit of small $\kappa_L$ resulting in a large value of $a(t)$ for a given $\mathcal{P}_{0}(t)$. Notice that we considered a double-sided cavity with left $\kappa_L$ and right $\kappa_R$ decay rates adding to the total loss rate $\kappa = \kappa_L + \kappa_R$. The coefficients $g^{(1)}_j$ and $g^{(2)}_j$ are the linear and quadratic per photon optomechanical coupling rates corresponding to the two situations depicted in Fig.~\ref{fig1}c and Fig.~\ref{fig1}d, respectively. While the cavity field amplitude inherently depends on the displacement of the mechanical mode, we will assume the unresolved sideband regime where this dependence is weak. Moreover, we are interested in the classical problem i.e. in simulating the proper set of actions that results in the shrinking of an initial large thermal distribution for the total energy of the oscillator. To this end we only consider the trivial dynamics of the cavity field classical amplitude $\alpha(t) = \langle a(t)\rangle$ which follows the driving field as $\dot{\alpha}(t) = -\kappa \alpha + \mathcal{E}(t)$. We can then reduce the dynamics of the system to
\begin{subequations}
\begin{align}
\label{OscEq.2a}
\dot{q}_j &= \omega_j p_j, \\
\label{OscEq.3a}
\dot{p}_j &= -\omega_jq_j - \gamma_{j} p_j + \xi_j - g^{(1)}_j|\alpha(t)|^2,\\
\label{OscEq.4a}
\dot{\alpha} &= -\kappa \alpha + \mathcal{E}(t),
\end{align}
\end{subequations}
which resemble Eqs.~\ref{FreeEq.1}a,b, where we can identify the action forces $F_j(t)=-g^{(1)}_j|\alpha(t)|^2$ (the cavity field $\alpha(t)$ playing the role of the action delivering the cooling momentum kicks to the mechanical oscillators). As noted before, as the actions are obtained from the same cavity field intensity, they only differ by the multiplicative $g^{(1)}_j$ factor. Notice also that this configuration strongly resembles a cold damping approach~\cite{Genes2008Ground, Steixner2005, Bushev2006}.\\
\indent In contrast, for a quadratic coupling Hamiltonian, the changes in the momentum are of a very different nature
\begin{eqnarray}
\label{eq4}
\dot{p}_j &=& -\left[\omega_j + 2g^{(2)}_j|\alpha(t)|^2\right]q_j - \gamma_j p_j + \xi_j,
\end{eqnarray}
as the cavity periodically modulates the oscillation frequencies of each mode. \\
\indent To provide the neural network feedback onto the motional dynamics, we use the inferred $q$ and $\dot{q}$ at a given time $t-\Delta t$ as input values for the neural network [see Fig.~\ref{fig1}]. The trained network then selects the appropriate action by choosing the value of $\mathcal{E}(t)$ (from a finite number of possible values) to be acted upon the system. The size of the time-step $\Delta t$ is chosen such that $1/\omega_{j} \gg \Delta t$ for all $j$ to minimize the error in the numerical integration. For a given drive amplitude, the set of actions on the different modes will be different according to the values of the optomechanical couplings (as they are proportional to $g^{(1)}_j|\alpha(t)|^2$ or $2g^{(2)}_j|\alpha(t)|^2$).We then use the Runge-Kutta fourth-order method (RK4) for the numerical integration of the dynamical system where we iteratively sum for each time step. Additionally, at each time step we inject the measured parameters of the dynamical system as input data into the nonlinear function formed by the neural network to predict on the action and thereby the momentum kick or cavity field strength suitable for the next time step, which is acquired from the output nodes (neurons) of the network.\\

\noindent \textbf{Reinforcement learning} --- The neural network provides a nonlinear function, that for some given input data, which harbor information about the oscillator states at a given time step $t$, predicts the correct action for the next time step $t+\Delta t$ that helps to reduce the overall energy of the dynamical system at later times. This function forms the neural network parametrized policy $\pi$. To obtain an optimal (or nearly optimal) policy we employ the technique of reinforcement learning~\cite{Sutton1998Reinforcement, Russel2018Modern} and in particular a policy gradient approach~\cite{Williams1992Simple}. Such a problem is in general referred to as a Markov decision process (MDP)\cite{Bellman1957Markovian} and described in detail in the Methods section. Here, the network acts as an agent that by observing parameters of the environment (resonator) improves its probabilistic policy that chooses the right actions $a_t$ at a given time $t$ to increase an overall reward $R = \sum_{t}R_{t}$ (full reward over a trajectory) that is connected to the reduction of the energy of the resonator modes. The actions are chosen from a finite set (of values of different amplitudes) and realized as momentum kicks or translated into frequency shifts. As an input to the network we feed information about the state of the environment given by $s_t = (q(t) ,\dot{q}(t))$. The network outputs the probabilities $\pi_{\theta}(a_t|s_t)$ for the actions $a_t$ that could be applied to the dynamical system. Here, the parameter $\theta$ encompasses all the weights and biases of the network. We take the action with the highest probability and apply it in the next iteration of Eq.~\ref{FreeEq.1} up to Eq.~\ref{OscEq.3a}. The probabilities $\pi_{\theta}(a_t|s_t)$ can be optimized with respect to an increased reward return $R_t$ by employing an update rule for the weights and biases of the neural network, following $\theta \leftarrow \theta + \Delta \theta$ and
\begin{eqnarray}
\label{ReinEq.1}
\Delta \theta_{j} &=& \eta \partial_{\theta_{j}} \mathbb{E}[R] = \eta \sum_{t}\mathbb{E}\left[(R-b)\partial_{\theta_{j}}(\ln \pi_{\theta}(a_t|s_t))\right],
\end{eqnarray}
where $\mathbb{E}$ is the expectation value over all state and action sequences (full trajectories), which here is approximated by averaging over a large enough set of oscillator trajectories (training batch) and their corresponding action sequences which we have obtained from the iterative summation of the dynamical equations (RK4) and from the predictions of the neural network at each time step for various randomly chosen initial conditions. The learning rate is given by the parameter $\eta$ and $b$ is a baseline to suppress fluctuations of the reward gradient \cite{Williams1992Simple,Weaver2001Optimal}. Here, the baseline is approximated by $b \approx b_n = (1/n-1)\sum_{i=1}^{n-1}\bar{R}^{(i)}$, where $\bar{R}^{(i)}$ is the average total Reward from the $i$'s learning epoch. Here, the training epoch is defined as the number of updates $\theta \leftarrow \theta + \Delta \theta$.\\
\indent The neural network which is represented by the array $\theta$ encompassing all weights and biases, consists of an input and output layer and two hidden layers whereby the number of input neurons depends on the number measured of variables of the system while the number of output neurons depends on the number of possible output actions, respectively (see Methods Tab.~\ref{Ap:Tb1}). The two hidden layers consists of up to $60$ to $100$ neurons each. The network is densely connected and we chose "relu" (rectified linear unit) as a nonlinear function acting on each neuron in the two hidden layers. The probabilities for each action given out by the output layer are obtained by using the "softmax" nonlinear function for the output neurons. From these probabilities the action is chosen by taking the neuron index with the highest probability value in the output.\\

\begin{figure*}[t]
\includegraphics[width=1.95\columnwidth]{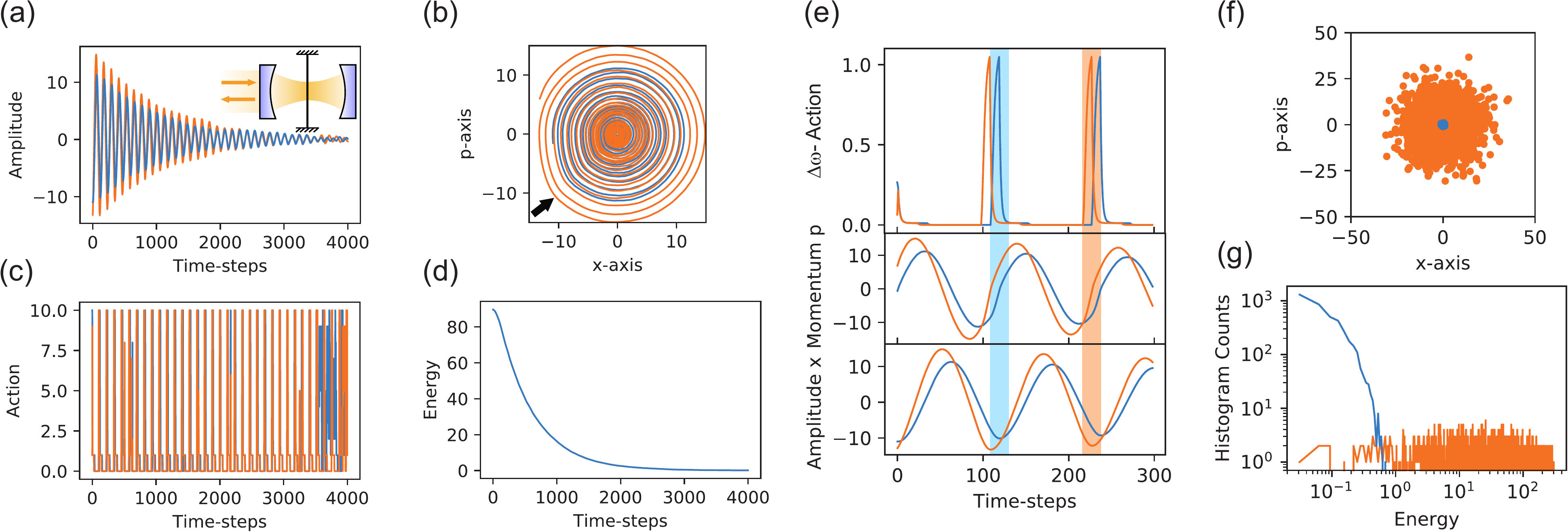}
\centering
\caption{\emph{Single mode parametric cooling}  (a) Time dynamics followed on two trajectories with initial conditions drawn from a Boltzmann distribution corresponding to an initial thermal state with average occupancy of $\bar{n} = 100$. The inset refers to the choice of cooling performed parametrically by modulation of the spring constant. (b) Corresponding phase space trajectories. The black arrow indicates where the force is applied. (c) Action sequences chosen by the network for each trajectory (d) Average energy for a thermal ensemble of trajectories exposed to the actions chosen by the network and rescaled to give the value of the occupation number of the harmonic oscillator. (e) Zoom-in into the time dynamics of the cavity modulated actions $\Delta\omega = 2g^{(2)}|\alpha(t)|^{2}$ and quadratures for two distinct trajectories. (f) Phase space comparison of initial (orange) and final (blue) distributions. The $4\times 10^{3}$ points of the distribution of final states are obtained by running each trajectory starting from its initial state under the actions of the policy for $2\times 10^{4}$ time steps. (g) Corresponding histogram of energy distribution in the initial and final states. The parameters are $\gamma = 4\times 10^{-5}\, \omega$, $g^{(2)} = 1\times 10^{-8}\,\omega$, $|\alpha|^{2} \approx 0.5\times 10^{7}$ and $\Delta t = 0.05\,\omega^{-1}$.}
\label{fig2}
\end{figure*}
\noindent \textbf{Single mode cooling} --- In a first step we numerically simulate the time dynamics of a single oscillating mode of frequency $\omega$ initially in a thermal distribution imposed by its coupling to an environment at some temperature $T$. This corresponds to the following distribution of energies
\begin{eqnarray}
\label{SinEq.1}
P(E) &=& Z^{-1}e^{-\beta E \hbar \omega},
\end{eqnarray}
with a partition function $Z \approx (\beta \hbar \omega)^{-1}$ and the total occupation number normalized energy $E = (q^2 + p^2)/2$. We then randomly pick an initial energy value from the thermal distribution
\begin{eqnarray}
\label{SinEq.2}
E_{0} &=& -(1/\beta \hbar \omega)\ln(1-s),
\end{eqnarray}
by picking a random number $s$ between zero and one. From the equipartition theorem we deduce $q(0)$ and $p(0)$ and train the neural network by recursively injecting sequences of $(q(t) ,\dot{q}(t))|_{[0\dots T]}$, obtained by applying the terms of Eqs.~\ref{OscEq.2a},\ref{OscEq.4a} and Eq.~\ref{eq4} recursively on the initial values, as training data into the network. A reward at each time step is only given when the action reduced the energy of the resonator at a given time step with respect to the previous time. The reward is defined by
\begin{eqnarray}
\label{SinEq.3}
R_t = (E_{0} - E_{t+\Delta t})\theta(E_{t}-E_{t+\Delta t}),
\end{eqnarray}
where $E_t$ is the total energy at time $t$. The reward gets larger when the energy separation between the current and initial energy increases therefore optimizing the effective cooling rate (see Methods Fig.~\ref{fig4}e). The application of the reward to the single mode cooling is exemplified for the quadratic coupling configuration illustrated in Fig.~\ref{fig1}d. The cooling dynamics is exemplified both as amplitude decreases Fig.~\ref{fig2}a and in phase space Fig.~\ref{fig2}b on two trajectories corresponding to two different initial states randomly picked from a thermal initial distribution with average occupancy $\bar{n}=100$. The action sequences of the network in Fig.~\ref{fig2}c show a periodic structure matching the frequency of the resonator mode, which is more visible in the zoom-in plot provided in Fig.~\ref{fig2}e. There one can follow the time dynamics of the applied action and the effect onto both the position and momentum, which in total for all trajectories results in the reduction of the average energy as presented in Fig.~\ref{fig2}d.\\
While the training of the neural network to produce an optimal policy is done with training batches of $80$ trajectories each with $4000$ time steps (already approaching a low energy steady state as presented in Fig.~\ref{fig2}a,b), we test the stability of the cooling policy by applying the trained network on a sample of thousands of trajectories with an extended time range of $20000$ time steps. These results of thousands of sample trajectories are presented as initial and final phase space distributions in Fig.~\ref{fig2}f and as a histogram of the energy distributions in Fig.~\ref{fig2}g. The injected thermal noise in all plots in Fig.~\ref{fig2} corresponds to a thermal occupation number of $\bar{n} \approx 100$ and a thermalization rate of $\gamma/\omega=4\times 10^{-5}$. The choice of the initial thermal state is however arbitrary and with equal computational power one can also describe the dynamics of oscillators initially populated with more than $10^5$ quanta.
These results show that the policy reduces the energy of the mechanical mode and converges at a steady state irrespective of the initial high energy state derived from the thermal distribution. Additionally, the larger data set does not show any divergent outliers suggesting that convergence has been reached.\\
\begin{figure*}[t]
\includegraphics[width=1.96\columnwidth]{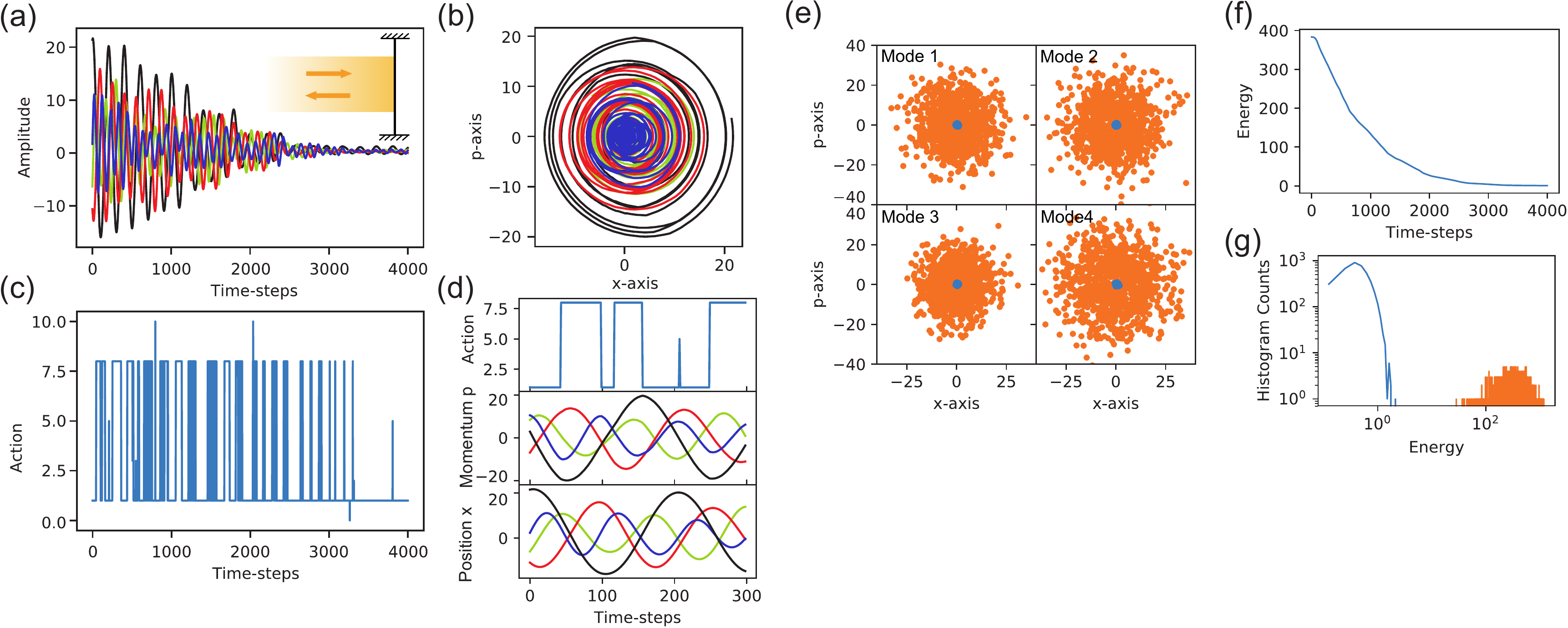}
\caption{\emph{Simultaneous cooling of 4 modes}. (a) Time dynamics of the oscillation amplitudes of four independent modes under the action of a collective force in the setup shown in the inset. (b) Corresponding time dynamics in phase space. (c) Neural network indicated sequence of actions leading to simultaneous cooling. (d) Magnification of the action and the momentum and position traces. (e) Reduction of the initial thermal distribution (orange) towards a low temperature distribution (blue) for each independent oscillation mode after $2\times 10^4$ time steps. (f) Corresponding decrease of the total average energy of all four modes. (g) Histogram of initial (orange) and final (blue) energy distribution for $4\times 10^3$ trajectories. The parameters for the simulation are given by $\omega_{2,3,4} = (0.8,\, 1.2,\, 0.6)\,\omega_1$, $\gamma_{1,2,3,4} = (4,\,3,\, 5,\, 2)\,\times 10^{-5} \omega_1$, where the multiplication factors for the action are $g_{1,2,3,4} = (0.3,\,0.2,\,0.4,\,0.3)$ to obtain $F_j$ and we have $\Delta t = 0.05\,\omega_1^{-1}$.}
\label{fig3}
\end{figure*}

\noindent \textbf{Simultaneous cooling of many modes} --- We display the generality of this approach by applying the network to find a strategy to cool up to four modes simultaneously. As in the case of the single mode cooling, we apply a single force on the mirror: this poses a challenge as a good cooling strategy for a given mode might actually lead to the heating of the other modes. In general, owing to this challenge, a simultaneous cooling strategy has an increased complexity in the choice of the action sequences, which leads to overall slower cooling rates. The same principles for cooling a single resonator are applied to cooling four modes subjected to the same actions by the network as presented in Fig.~\ref{fig3}. Here, we use the setup configuration presented in Fig.~\ref{fig1}b. While the four modes have different frequencies $\omega_j$ and coupling strengths $g_j$ they are subjected to the same time sequence of actions delivered by intensity variations of an impinging laser beam in free space or via the field intensity $|\alpha(t)|^2$ in the cavity.

The input to the network is given by $s_t = (Q(t), \dot{Q}(t))$ with $Q(t) = \sum_{j} q_{j}(t)$ and $\dot{Q}(t) = \sum_{j} \omega_jp_{j}(t)$ as a collective position coordinate and its derivative. The quantities can be obtained for example from an interferometer that is sensitive to the fluctuations on the membrane (see Fig.~\ref{fig1}e) or via homodyne detection. The derivation of the initial condition is described in the Methods section.\\
Here, the agent needs to find a strategy that simultaneously cools the center of mass motion as well as all of the relative mode dynamics. As an example the trajectories of the four modes are presented in Fig.~\ref{fig3}a,b, where the cooling results from the corresponding actions presented in Fig.~\ref{fig3}c with magnified view shown in Fig.~\ref{fig3}d. In contrast to the action sequence imposed on the trajectory of a single resonator presented in Fig.~\ref{fig2}c, which basically shows a periodic signal matching the frequency of the oscillator, here we find a more complex signal with a quasi periodic pattern. The change of the average energy of the four resonators as a function of time is presented in Fig.~\ref{fig3}f. In Fig.~\ref{fig3}e the initial values obtained from a Boltzmann distribution and final phase space values of a thousand trajectories for all oscillators are presented. A histogram of the sum of their individual energies is given in Fig.~\ref{fig3}g. These results show that all four resonators can be simultaneously cooled down to lower temperatures, that differ by orders of magnitude from their initial values and thereby exemplify the strength of this adaptive approach.\\

\noindent \textbf{Conclusions} --- We have shown results of numerical simulations for the simultaneous cooling of a few degrees of freedom of a vibrating mechanical resonator. The feedback action has been realized via the reinforcement learning technique implemented on a neural network. There is a variety of other optimization methods that could obtain similar results. For example, stochastic optimization methods such as hill climbing, random walks or genetic algorithms.\cite{Russel2018Modern} It has been recently shown that evolution strategies (ES) offer a similar performance and efficiency as RL.\cite{salimans2017evolution} We have selected RL and especially the policy gradient method to approach this problem due to its efficiency when a large continuous or quasi continuous set of states is present.\cite{dutta2018Reinforcement}\\ Simultaneous cooling of a few modes indicates the possibility of partial or full refrigeration of mechanical resonators via optical control. It is remarkable that the network can perform efficient cooling of many modes while only being fed information of a time-evolving collective displacement quadrature. This is owed to the fact that the designed strategy optimizes single mode cooling at every instance in time while keeping the heating of all other modes to small values. The described procedure works both inside and outside optical cavities and both in linear or nonlinear regimes therefore being easily adaptable to new systems. The technique could be easily extended to cool a number of oscillators or a number of particles trapped inside optical cavities or with tweezers. While the present treatment considers classical stochastic dynamics, a full quantum theory of neural network aided cooling will be tackled in the future that might also lead towards feedback production of squeezed or squashed states. In this regard, for both single and many oscillation modes, we plan to analyze the efficiency of neural network cooling in comparison with standard cold-damping optical cooling. It is expected that a Fourier analysis of the action function indicated by the network could hint towards feedback implementations that could surpass existing techniques, especially at the level of many degrees of freedom.\\

\noindent\textbf{Acknowledgements}

We acknowledge financial support from the Max Planck Society. We acknowledge fruitful discussions with Michael Reitz. We are thankful for the comprehensive lecture notes on machine learning for physicists presented by Florian Marquardt at the University of Erlangen-Nuremberg~\cite{Marquardt}.

\bibliography{bibfileReinforcement}

\onecolumngrid
\newpage
\appendix

\renewcommand{\theequation}{A.\arabic{equation}}
\renewcommand\thefigure{A.\arabic{figure}}
\setcounter{equation}{0}
\setcounter{figure}{0}

\newpage
\section{Initial conditions}
We assume an initial Boltzmann distribution of energies $P(E_1, \dots, E_n) = Z^{-1}e^{-\beta(E_1+\dots+ E_n)}$ where $Z = 1/\beta^n$, from which we extract the initial conditions by integration
\begin{eqnarray}
\label{Ap:Eq.1}
\nonumber
\int^{\tilde{E}_1}_{0}\dots \int^{\tilde{E}_n}_{0} dE_{1}\dots dE_{n}P(E_1, \dots, E_n)= \prod _{j=1}^{n} \left[1 - e^{-\beta\tilde{E}_{j}} \right] =s,
\end{eqnarray}
where $s$ is a value between $0$ and $1$. We set $b_j = \left[1 - e^{-\beta\tilde{E}_{j}} \right]$ and define that $b_j \in [0,1]$ for $0 \leq j \leq n$ which results in
\begin{eqnarray}
\label{Ap:Eq2}
\tilde{E}_{j} &=& -\frac{1}{\beta}\ln(1-b_j).
\end{eqnarray}
Since $p_{j}(0)^{2} + q_{j}(0)^{2} = 2\tilde{E}_{j}/\hbar\omega_j$ we obtain $p_j$ and $q_j$ from
\begin{subequations}
\begin{align}
\label{Ap:Eq3}
q_{j}(0) &= \sqrt{\frac{2\tilde{E}_{j}}{\hbar\omega_j}}\cos(2\pi\phi_{j}),\\
p_{j}(0) &= \sqrt{\frac{2\tilde{E}_{j}}{\hbar\omega_j}}\sin(2\pi\phi_{j}),
\end{align}
\end{subequations}
where $\phi_{j}$ is a random number between $0$ and $1$ for all $j\in {1,\dots, n}$.

\section{Thermalization dynamics}
\begin{figure}[t]
\centering
\includegraphics[width=0.8\columnwidth]{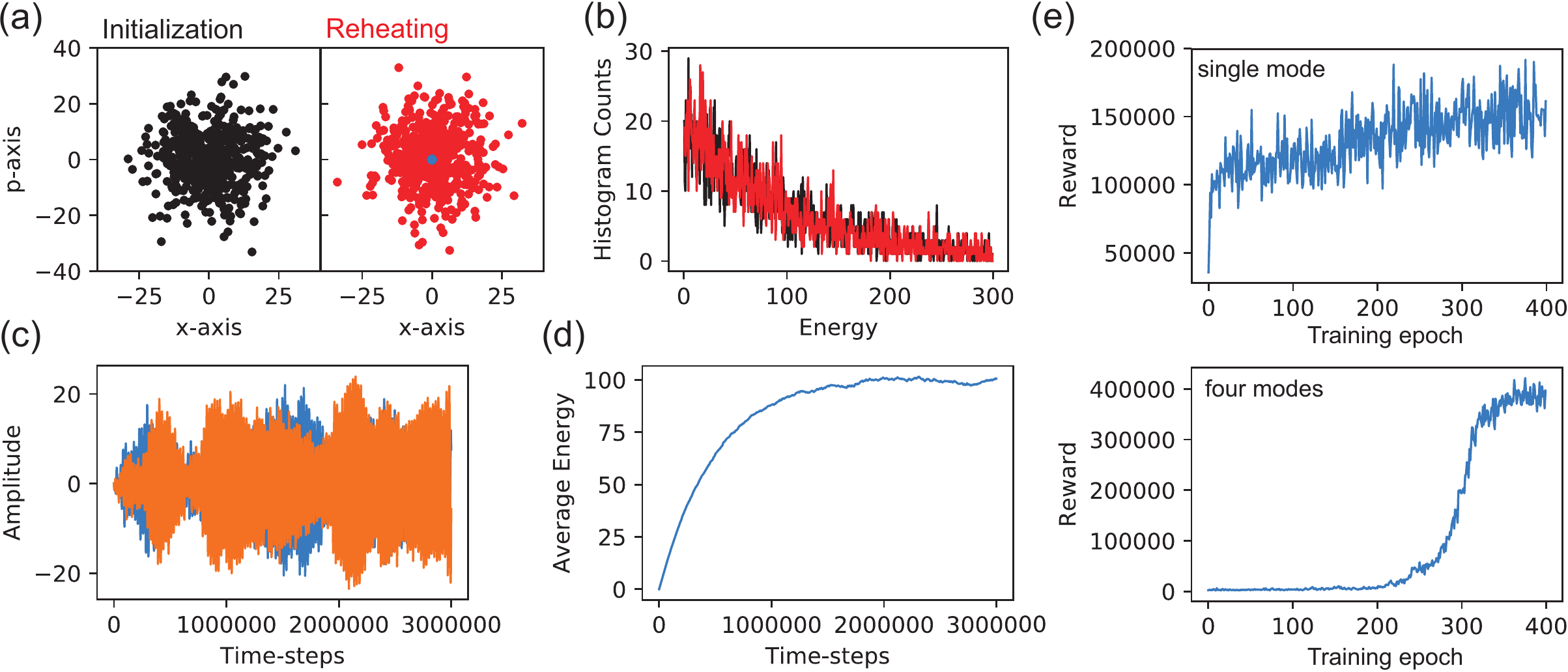}
\caption{\emph{Thermalization dynamics of a single mode of vibration}. (a) Phase space coordinates from initialization with a Boltzmann distribution with $\bar{n} = 100$ (black points) and obtained from the evolution equation containing thermal noise. An initial distribution around the phase space origin (blue points) is evolved up to the steady state (red points). In (b) the corresponding energy distributions following the same color coding are presented. (c) Two example trajectories with starting energies close to zero are following the evolution guided by thermal noise. In (d) the average energy is presented as a function of time. (e) The learning progress for the single and four-mode cooling presented in the main text. Here, the reward increases steadily with each training epoch until reaching a saturation.}
\label{fig4}
\end{figure}

Let us describe the numerical procedure for simulating the action of a thermal environment onto the state of the mechanical resonator. We consider the equations of motion
\begin{subequations}
\begin{align}
\label{Ap:Eq4}
\dot{q} &= \omega p,\\
\dot{p} &= -\omega q - \gamma p + \xi,
\end{align}
\end{subequations}
which we rewrite as equations of differential forms
\begin{subequations}\label{Ap:Eq5}
\begin{align}
dq &= \omega p dt, \\
\label{Ap:Eq5a}
dp &= -\omega q dt - \gamma p dt + \sqrt{(2\bar{n}+1)\gamma}dW(t),
\end{align}
\end{subequations}
where the noise $dW(t)$ is included as a Wiener process. Numerically, this can be realized by $\Delta W(t) \propto \sqrt{\Delta t}N(0,1)$, where $N(0,1)$ describes a normally distributed random variable of unit variance - consequently the Wiener increment is normally distributed with a variance equal to the numerical time increment $\Delta t$. As a numerical check we simulate the thermalization of an initially cold mechanical mode under the action of an environment with rate $\gamma=4 \times 10^{-5}\, \omega$ and at an effective occupancy $\bar{n}=100$. The analytical Boltzmann distribution for this occupancy is shown in Fig.~\ref{fig4}a as black dots while the final state obtained from the numerical integration is represented by the red dots (blue dots in the middle are the initial state). In Fig.~\ref{fig4}b the agreement between the numerical simulation and the Boltzmann distribution is illustrated as a histogram of energy states. The time evolution is shown in Fig.~\ref{fig4}c as dynamics for the position and in Fig.~\ref{fig4}d for the total energy showing it approaching $\bar{n}=100$ in the long time limit.\\
Additionally, in Fig.~\ref{fig4}e we present the learning process for the results presented in Fig.~\ref{fig2} for a single mode and  for four modes as presented in Fig.~\ref{fig3}, where the increase in the average reward over $400$ epochs each with batches of $80$ trajectories is shown.

\section{Markov decision process}
The reinforcement learning procedure presented above can be fully described by a discrete time stochastic control process. Since in this process the agent selects an action based on the stochastic policy $\pi_{\theta}(a|s)$ which is only dependent on the current observation of the given state $s$, the problem is described by a Markov decision process (MDP)\cite{Bellman1957Markovian}. Formally, a Markov decision process is a 4-tuple $(S, A, P, R)$, where $S$ is the set of states, $A$ forms the set of possible actions, $P: S\times A \times S \rightarrow [0,1]$ is the transition function between states and $R:S\times A \times S \rightarrow \mathcal{R}$ is the reward function as described above with $\mathcal{R} = [R_{\text{min}},R_{\text{max}}] \subset \mathbb{R}$ being the continuous set of possible rewards.\\

In the case of a single resonator mode the state space is given by $S = \{s= (q,p)| \; p = \omega^{-1}\dot{q}, \; q,p \in \mathbb{R}\}$ while the action space is the finite set $A = \{0,1,\dots,10\}$ which allows the agent to choose between ten different force strength. The transition function $P(s_{t+\Delta t}|a_{t},s_{t})$ giving the probability for moving to the state $s_{t+\Delta t}$ from $s_{t}$ under the action $a_{t}$ can be obtained from the equations of motion $\dot{s} = Ms + \tilde{\xi} + \tilde{a}$ with $\tilde{\xi} = (0,\xi)^{\top}$ being the noise and $\tilde{a} = (0, F(a))^{\top}$ the force term. In the case the random noise contribution $\xi$ is zero the transition function is deterministic and $P(s_{t+\Delta t}|a_{t},s_{t}) = 1$ for $s_{t+\Delta t} = \tilde{M}s_{t} + \tilde{a}_{t}\Delta t$ with $\tilde{M} = 1 + M\Delta t$ and zero otherwise.
The reward function is given by the expression defined above $R_{t} = R(s_{t+\Delta t},a_{t},s_{t}) = (E_{0}-E_{t+\Delta t})\theta(E_{t}-E_{t+\Delta t})$.\\
For multiple resonator modes where we observe $s_t = (Q(t),\dot{Q}(t)))$ with $Q(t) = \sum_{j}q_{j}(t)$ and $\dot{Q}(t) = \sum_{j}\omega_{j} p_{j}(t)$, we only obtain partial information of the state which originally is described by the phase space vector $(q_{1},\dots,q_{n},p_{1},\dots, p_{n})$. Here we need a generalization of an MDP which is given by a partially observable Markov decision process (POMDP), where the agent cannot observe the full state. Here, we additionally have the sets $\Omega$ which describes the set of observations and $O$ describing the set of conditional observation probabilities.\\
A pseudo code to implement reinforcement learning (RL) is presented in the following:
\begin{eqnarray}
\label{Ap:Eq.6}
\nonumber
& & \text{initialize}\; \theta \\\nonumber
& & \text{for iteration}\; = 1,2,\dots,N \; \text{epoch}\; \text{do}\\\nonumber
& & \hspace{.5in} \text{initialize training batch of initial states} \; s \\\nonumber
& & \hspace{.5in} \text{for each time step} \; t \; \text{do} \\\nonumber
& & \hspace{1.0in} a_{t} = \text{arg max}_{a' \in A} \pi_{\theta}(a',s_{t}) \quad \text{sample action for the full batch} \\\nonumber
& & \hspace{1.0in} s_{t+\Delta t} = \tilde{M}s_{t} + d\tilde{W} + \tilde{a}_{t}\Delta t \quad \text{run dynamics: for the full batch}\\\nonumber
& & \hspace{.5in} \text{end for}\\\nonumber
& & \hspace{.5in} R = \sum_{t} R_{t} \quad \text{total reward for each trajectory of the batch}\\\nonumber
& & \hspace{.5in} \theta \leftarrow \theta + \eta\nabla_{\theta}\mathbb{E}(R) \quad \text{update network parameters}\\\nonumber
& & \text{end for}
\end{eqnarray}
\subsection{Network parameters}
The simulations were run on a standard Laptop computer (CPU, Intel Core i$7-5500$U @ $2.40\,$GHz). We use the Keras package for Python and the Theano framework~\cite{AlRfou2016Theano} to realize the neural network and the reinforcement learning procedure. The network and training parameters are presented in Tab.~\ref{Ap:Tb1}.
\begin{table}[h]
\centering
\begin{tabular}{ l || r }
\hline
 & Network parameters \\ \hline \hline
Training batch size & 80 \\
Learning rate $\eta$ (Adam optimizer) & 0.00008 (single oscillator) \\
 & 0.0006 (four oscillators) \\
Neurons per layer & $(2, 60, 60, 11)$ (single oscillator) \\
& $(2, 100, 100, 11)$ (four oscillators) \\
Reward scale & 1.0 \\ \hline \hline
\end{tabular}
\caption{\emph{List of neural network and training parameters.}}
\label{Ap:Tb1}
\end{table}

\end{document}